# Donut visualizations for network-level and regional-level overview of Spatial Social Networks

Dipto Sarkar
Department of Geography
University College Cork
Cork, Ireland
dipto.sarkar@ucc.ie

Piyush Yadav
Lero-SFI Irish Software Research Centre, Data Science Institute
*National University of Ireland Galway (NUI Galway)*
Galway, Ireland
piyush.yadav@lero.ie

*Abstract*—Spatial Social Networks (SSN) build on the node and edge structure used in Social Network Analysis (SNA) by incorporating spatial information. Thus, SSNs include both topological and spatial data. The geographic embedding of the nodes makes it impossible to move the nodes freely, rendering standard topological algorithms (e.g. force layout algorithms) used in SNA ineffective to visualize SSN sociograms. We propose a new visualization technique for SSNs that utilize the spatial and social information to provide information about the orientation and scale of connections. The donut visualization can be used to summarize the entire network or can be used on a part of the network. We demonstrate the effectiveness of the donut visualization on two standard SSNs used in literature.

## I. Introduction

Spatial Social Networks (SSN) incorporate spatial information in the node and edge structure of social networks to capture the spatial embedding of the network. In its simplest form, SSNs incorporate spatial information as a spatial (x, y) nodal attribute [1]. Even though SSNs are based on social networks, standard algorithms used in Social Network Analysis for visually representing node-and-edge structures as sociograms are of limited use in case of SSNs. Layout algorithms (e.g. force-directed layouts) rely solely on the topological structure of the network and move the nodes freely to create aesthetically pleasing andinterpretable sociograms. As long as the topology (node to edge connections) is preserved, the positions of the nodes do not matter. In the case of SSN, the nodes are anchored in geographic space using the x, y location information. The geographic embedding restricts the ability to move the nodes. In these location anchored sociograms, the edges are drawn as graphical artifacts and do not have explicit spatiality themselves [2].

On the other hand, using the geo-located nodes along with base maps can provide additional information about the situation of the network, such as resources available to the nodes, barriers and promoters of social relationships, and spatial orientation of connections. Thus, there is a conundrum when visualizing SSNs as sociograms. Either the user must forgo the map and use topology-preserving sociograms, trying to portray as much spatial information as possible using visual elements like color and shape of the nodes. Or, the user prioritizes the spatial aspect of the network and draws the nodes and edges on a base map, which can lead to hairball like uninterpretable structures as the size of the network increases. Hence, there are significant challenges for visualizing SSNs and creative techniques are required to highlight different socio-spatial aspects at different scales [1, 3–7].

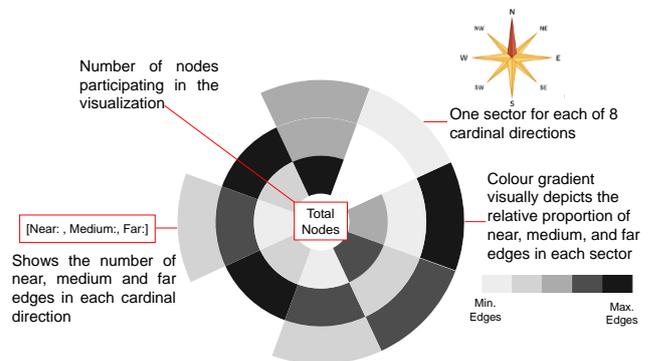

*Figure 1: Concept diagram showing how the donut visualization depicts an SSN*

We propose a polar stacked column chart based rose diagram like SSN visualization technique similar to [8] that utilizes the spatial as well as the social information in SSN and breaks away from the traditional node-and-edge sociogram representation. The Donut Plot provides information about the overall socio-spatial properties of the entire or part of the network in focus. The visualization explicitly provides information about scale, both in terms of the number of connections (topological) and spatial distribution of connections (spatial), as well as the spatial orientation of links at the network-level and regional-level.

## II. Concept, Implementation and Data

### A. Concept

Figure 1 shows a concept donut diagram. Radial diagrams focus on the centroid and the sectors to convey information [9]. The visualization consists of 8 segments corresponding to cardinal directions. The center of the donut diagram is positioned at the centroid of the layout view on which the social network is overlaid. The sectors are used to portray the directionality of connection in the network. Each segment is further broken down into three wedges to show the number of near, medium, and far connections in that direction. The color of the wedges provides a visual depiction of the number of connections. The number at the center depicts the number of nodes participating in the visualization.

The map containing the social network can be moved and zoomed in when creating the visualization. Only the nodes visible in the current viewport is used to create the donut chart. The number at the center also updates to reflect the data visible in the viewport. The donut chart can provide regional-level information if the visualization is generated after focusing on only the part of the network which is of interest. Hence, the donut chart is a flexible tool able to provide a visual summary of either the entirety or spatially contiguous part of the network.

Even while zoomed-out, the donut chart provides perspectives on the network across several scales. The sectors and wedges break down the connections according to directions and distance to communicate the structure of the network across scales. Thus, the donut chart can 'probe' the network, ensuring that even zoomed out network-level views to not completely suppress local and regional trends [10].

The visualization is capable of handling both directed and undirected networks. In the case of directed networks, the ordered pair of nodes to which an edge is connected is used to determine the directionality of connection. In the case of undirected networks, the nodes participating in the edge are unordered and it is not possible to determine the origin and destination. Hence, each edge is counted twice, once for going from node u to node v and once for going from node v to u.

**Algorithm 1**
**Data:** Spatial Network Layers
**Output:** Aggregated Spatial Network Donut Graph
**Procedure:**
$\{Layer_1, Layer_2, \ldots, Layer_n\} \leftarrow LoadSpatialNetwork()$
$ActiveLayer \leftarrow getActiveLayer(Layer_1, \ldots, Layer_n)$
$LayerExtent \leftarrow getLayerExtent(ActiveLayer)$
$\{P_{east}, P_{south-east} \ldots P_{north-east}\} \leftarrow getdirectedPolygon(LayerExtent)$
$Agg.GraphData \leftarrow createEmptyGraphData()$

**for** *each* layer in $\{Layer_1, Layer_2, \ldots, Layer_n\}$ **do**
  $NodeLayers \leftarrow identifyNodeLayers(layer)$
  $EdgeLayers \leftarrow identifyEdgeLayers(layer)$
**end**

**for** *each* $layer_{node}$ in $NodeLayers$ **do**
  $nodedirection \leftarrow getNodeDirections(layer_{node})$
  $nodefeatures \leftarrow getNodeFeatures(layer_{node})$
  $Agg.GraphData \leftarrow Agg.GraphData \cup$
  $(updateAgg.GraphData(nodedirection, nodefeatures))$
**end**

**for** *each* $layer_{node}$ in $EdgeLayers$ **do**
  $\{src.node, dest.node, length\} \leftarrow getEdgeFeatures(layer_{edge})$
  $length_{norm} \leftarrow NormalizeEdgeLength(layer_{edge})$
  $length_{bucket} \leftarrow identifyLengthBucket$
            $(\{Near, Far, Medium\}, length_{norm})$
  $Agg.GraphData \leftarrow Agg.GraphData \cup$
  $(updateAgg.GraphData(src.node, dest.node, length_{norm}, ))$
**end**
$createAgg.SpatialNetworkGraph(Agg.GraphData)$

*B. Implementation*

The visualization is operationalized as a QGIS plugin implemented in Python 3. The plugin loads the network via QgsProject instance over QGIS canvas. The plugin reads all the visible/active layers to create the donut visualization.

Algorithm 1 defines the step by step process to create the aggregated graph of the spatial network. Initially, the plugin identifies the layer extent and create a directed polygon $\{P_{east}, P_{south-east}, \ldots P_{north-east}\}$ for all the eight directions. Later it identifies the direction of the node using the directed polygon as a reference. The plugin fetches the node and edge features and continuously update the aggregated graph data ($Ag.GraphData$). The system gets all the edge length and normalizes it between 0 and 1. Later it allocates the edges into three buckets- near, medium and far based on normalized length. The bucketing threshold is configurable and can be changed by the user according to his length definition. For the current implementation, the threshold is set as: 1) Near <=0.35, 2) Medium (>= 0.36 and <= 0.60) and 3) Far (>=0.61). Finally, the system creates the aggregated view of the spatial network using the updated information from $Ag.GraphData$. The plot is saved and viewed using matplotlib python library.

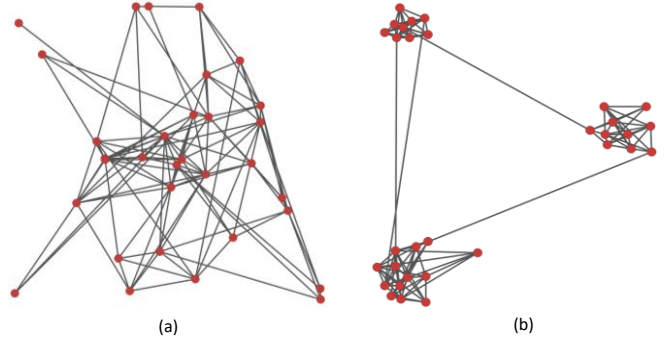

*Figure 2. Synthetic SSNs used for demonstrating efficacy. (a) The nodes are distributed in a Poisson manner. The probability of a link connecting two nodes decreases with distance. (b) The nodes are distributed in a randomly in three clusters. The probability of a link connecting two nodes decreases with distance.*

*C. Data*

We use the synthetic undirected networks used by [5] to demonstrate the efficacy of our techniques. Specifically, the synthetic networks (Figure 2) are idealistic representations that mimic the properties found in real-world network and provide a standard test data. In the synthetic networks, nodes are located randomly either in distributed (Figure 2a) or clustered (Figure 2b) manner in arbitrary Euclidean space. In both cases, the probability of connections decreasing with distance. Both the networks are depicted using location anchored sociograms, thus retaining the orientation and distance of nodes and edges with respect to each other.

### III. RESULTS

Figure 3a, the donut diagram corresponding to the sociogram Figure 2a highlights that nodes are present in all directions and shows the distance decay clearly through the color gradient going from dark to light as one moves from the center to the periphery. Even though the network was created with a Poisson node distribution, there are slightly higher number of nodes and edges in the South-East part. In the North-West region of the network, there are no near edges as all the connections are long or medium range and connect to nodes in the other sector. A similar, but slightly less stark version of this phenomenon is seen in the South-West sector too.

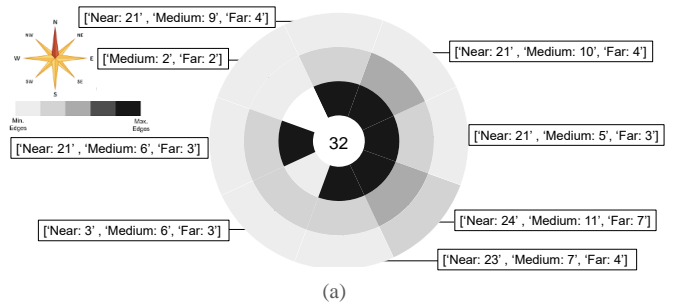

(a)

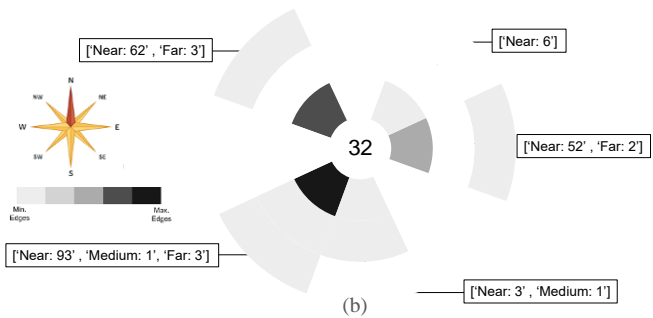

*Figure 3: Donut diagrams corresponding to (a) Figure 2a and (b) Figure 2b*

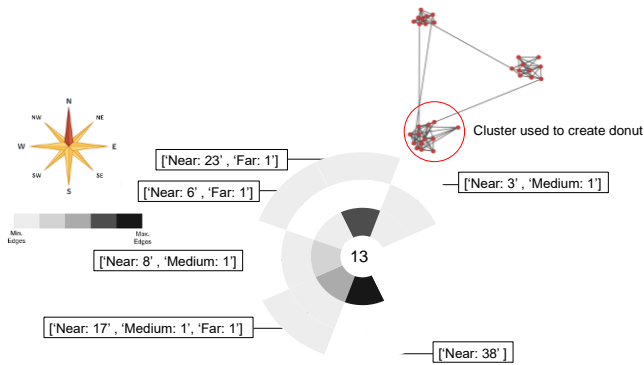

*Figure 4: Shows the donut diagram generated by focusing only on the highlighted part of the clustered random network shown in Figure 2b.*

Figure 3 (b), the donut diagram corresponding to the sociogram Figure 2 (b) is much sparser compared to Figure 3a even though the networks have a similar number of nodes and edges. This highlights the difference in the socio-spatial structure of the two networks. The three prominent spokes with a color gradient going from dark to light (namely North, North-West, and South-West direction) imply that the network has three clusters of nodes with connections predominantly within the same cluster. The slightly higher number of far connections in the North-West and South-West spokes alludes to more far relationships being maintained from those sectors. Only the South-West sector, maintains medium distance connections due to the relations maintained to the slightly remote node present to the East of the South-West cluster. Figure 4 shows the donut visualization created by zooming in on the 13 nodes in the South-West cluster only. The three long-distance connections are the ones maintained by nodes belonging to this cluster to the other two clusters. The location of the nodes that maintain the long distance connections are also evident. Apart from the far connections, most of the connections in this cluster are local with only a few medium distance connections. Note that only some (three out of five) of the connections to the remote node of the cluster are classified as medium distance.

## IV. CONCLUSION

Despite the restrictions put on embedding nodes in Euclidean space, most SSN visualizations still rely on sociograms. The donut chart utilizes the spatial and topological information but is liberated from their constraints as it not situated in either Euclidean or network space. The visualization provides an overview of the structure of the SSN in terms of overall directionality and distance distribution of connections. Moreover, the donut chart can provide these insights at both network-level and regional-level. Thus, the donut chart contributes an alternative visualization technique to sociograms for SSNs capable of communicating the spatial distribution of social connections. However, the donut chart is not meant to be used independent of the sociogram. The sociogram and the donut chart should work in tandem. In real-world networks, the donut chart presents an overall socio-spatial structure in terms of the number of nodes, directionality, and distance of connections in the network. The sociogram can then be used to look at the individual regions and connections of interest. For example, in Figure 4, the South-West sector could be investigated in depth to see the exact pattern of connections there that make that sector stand out in the visualization. In conclusion, this paper highlights the dichotomy in SSNs visualizations in incorporating both social and spatial aspects and presents ways forward with alternate visualization techniques that work in tandem with traditional sociograms to glean information from the network.

REFERENCES

[1] D. Sarkar, R. Sieber, and R. Sengupta, "GIScience Considerations in Spatial Social Networks," in *Lecture Notes in Computer Science*, 2016, vol. 1, pp. 85–98, doi: 10.1007/978-3-319-45738-3_6.

[2] C. Andris, "Integrating social network data into GISystems," *International Journal of Geographical Information Science*, vol. 8816, no. March, pp. 1–23, Mar. 2016, doi: 10.1080/13658816.2016.1153103.

[3] M. Batty, "The Geography of Scientific Citation," *Environment and Planning A: Economy and Space*, vol. 35, no. 5, pp. 761–765, 2003.

[4] A. Comber, M. Batty, and C. Brunsdon, "Exploring the Geography of Communities in Social Networks," in *Proceedings of the GIS Research UK 20th Annual Conference*, Lancaster, 2012, pp. 33–37, Accessed: Oct. 05, 2014. [Online].

[5] D. Sarkar, C. Andris, C. A. Chapman, and R. Sengupta, "Metrics for characterizing network structure and node importance in Spatial Social Networks," *International Journal of Geographical Information Science*, vol. 33, no. 5, pp. 1017–1039, May 2019, doi: 10.1080/13658816.2019.1567736.

[6] C. Andris and D. Sarkar, "Methods for the Geographic Representation of Interpersonal Relationships and Social Life," in *Abstracts of the ICA*, 2019, vol. 1, no. 1995, pp. 1–2, doi: 10.5194/ica-abs-1-11-2019.

[7] U. Demšar, J. Reades, E. Manley, and M. Batty, "Revisiting the Past: Replicating Fifty-Year-Old Flow Analysis Using Contemporary Taxi Flow Data," *Annals of the American Association of Geographers*, vol. 108, no. 3, pp. 811–828, May 2018, doi: 10/gc4n42.

[8] G. Andrienko, N. Andrienko, G. Fuchs, and J. Wood, "Revealing Patterns and Trends of Mass Mobility Through Spatial and Temporal Abstraction of Origin-Destination Movement Data," *IEEE Transactions on Visualization and Computer Graphics*, vol. 23, no. 9, pp. 2120–2136, Sep. 2017, doi: 10/gfw9kx.

[9] G. M. Draper, Y. Livnat, and R. F. Riesenfeld, "A Survey of Radial Methods for Information Visualization," *IEEE Transactions on Visualization and Computer Graphics*, vol. 15, no. 5, pp. 759–776, Sep. 2009, doi: 10/cz46vt.

[10] T. Butkiewicz, W. Dou, Z. Wartell, W. Ribarsky, and R. Chang, "Multi-Focused Geospatial Analysis Using Probes," *IEEE Transactions on Visualization and Computer Graphics*, vol. 14, no. 6, pp. 1165–1172, Nov. 2008, doi: 10.1109/TVCG.2008.149.